\documentstyle[aps]{revtex}

\input epsf
\def\jepsfbox#1{\typeout{#1} \epsfbox{#1}}
\def\plotonesc#1#2{\begin{center} \leavevmode
\epsfxsize=#2\columnwidth \jepsfbox{#1} \end{center}}

\def\jcite#1#2{#1 \cite{#2}}
\def\rarrow{\rightarrow}
\def\etal{{\it et al.\ }}
\def\eg{{\it e.g.~}}
\def\ie{{\it i.e.~}}

\def\rmmat#1{{\hbox{\rm #1}}}
\def\rmscr#1{\rmmat{\scriptsize #1}}
\newcommand{\be}{\begin{equation}}
\newcommand{\ee}{\end{equation}}
\newcommand{\ba}{\begin{eqnarray}}
\newcommand{\ea}{\end{eqnarray}}
\def\p{\partial}
\def\d{{\rm d}}

\def\dd#1#2{\frac{\d #1}{\d #2}}
\def\pp#1#2{\frac{\p #1}{\p #2}}
\def\figref#1{Fig.~\ref{fig:#1}}
\def\eqref#1{Eq.~\ref{eq:#1}}
\begin{document}
\draft
\newcommand{\bfi}{{\bf B}}
\newcommand{\hfi}{{\bf H}}
\newcommand{\efi}{{\bf E}}
\newcommand{\lag}{{\cal L}}
\newcommand{\dLIII}{{\frac{\partial^3 \lag}{\partial I^3}}}
\newcommand{\dLII}{{\frac{\partial^2 \lag}{\partial I^2}}}
\newcommand{\dLI}{{\frac{\partial \lag}{\partial I}}}
\newcommand{\dLKKK}{{\frac{\partial^3 \lag}{\partial K^3}}}
\newcommand{\dLKK}{{\frac{\partial^2 \lag}{\partial K^2}}}
\newcommand{\dLK}{{\frac{\partial \lag}{\partial K}}}
\newcommand{\dLIK}{{\frac{\partial^2 \lag}{\partial I \partial K}}}
\title{QED One-loop Corrections to a Macroscopic Magnetic Dipole}

\author{Jeremy S. Heyl \and Lars Hernquist}
\address{Lick Observatory, 
University of California, Santa Cruz, California 95064, USA}
\maketitle
\begin{abstract}
We consider the field equations of a static magnetic field including
one-loop QED corrections, and calculate the corrections to the field
of a magnetic dipole.
\end{abstract}

\section{Introduction}

The one-loop corrections of quantum electrodynamics introduce
nonlinearities in the equations of the electromagnetic field.  These
corrections manifest themselves through an index of refraction,
electric permittivity and magnetic permeability tensors which 
are a function of field strength\cite{Tsai75,Heyl97b}.  The vacuum
responds to an applied field like a nonlinear paramagnetic substance
\cite{Miel88}.   Using an analytic expression for the effective
Lagrangian of QED to one-loop order \cite{Heyl97a}, we derive
the magnetic magnetic permeability tensor as a function of the applied
magnetic field and calculate the one-loop corrections to the field of
a macroscopic magnetic dipole.

\section{The Effects of Non-linearity on the Field Equations}

The effective Lagrangian of QED at one-loop order consists of the sum
of a linear and a non-linear term
\be
\lag = \lag_0 + \lag_1.
\label{eq:lagdef}
\ee
Both terms of the Lagrangian can be written in terms of the Lorentz
invariants,
\be
I = F_{\mu\nu} F^{\mu\nu} = 2 \left ( |\bfi|^2 - |\efi|^2 \right )
\label{eq:Idef}
\ee
and
\be
K = - \left (\frac{1}{2} \epsilon^{\lambda\rho\mu\nu}
F_{\lambda\rho} F_{\mu\nu} \right)^2 =
	- (4 \efi \cdot \bfi )^2.
\label{eq:Kdef}
\ee
following Heisenberg and Euler \cite{Heis36}.  The Greek indices 
count over space and time components ($0,1,2,3$).

Since we are interested in the static properties of a magnetic
field, we can take $K=0$.
If we apply the Euler-Lagrange condition to extremize the action, we
obtain,
\ba
\nabla \times \hfi = 0 \\
\hfi = -4 \dLI \bfi
\ea
where the factor of $-4$ is inserted for later convenience.

Since $\nabla \times \hfi = 0$, we take $\hfi=-\nabla \phi$,
where $\phi$ is the magnetic scalar potential.  Furthermore,
the field $\bfi$ is derived from a vector potential
(\ie $\bfi=\nabla \times {\bf A}$) and so we also
have,
\be
\nabla \cdot \bfi = 0.
\ee
If the relationship between $\hfi$ and $\bfi$ were linear, this
equation would be satisfied by $\nabla^2 \phi=0$.  However, we will
assume a small non-linearity between the two fields,
\be
-4 \dLI = \mu'_0 + \mu'_1(B^2)
\ee
where $\mu'_0$ is a constant and $\mu'_1$ a function such that $\mu'_1(B^2)
\ll \mu'_0$.

We can invert the relationship between the two fields to first order
\be
\bfi = \left ( \mu_0 + \mu_1(H^2) \right ) \hfi
\ee
Now we recast the field equation with the magnetic potential
\be
\nabla \cdot \left [ \left ( \mu_0 + \mu_1(H^2) \right ) \nabla \phi
\right ] = 0,
\ee
write $\phi=\phi_0+\phi_1$, and solve the equation order by order
\def\MU11{\mu_1^{(1)}}
\def\RHOeff{\rho_\rmscr{eff}}
\ba
\nabla^2 \phi_0 &=& 0  \\
\nabla^2 \phi_1 &=& \RHOeff =
- \nabla \cdot \left [ \frac{\mu_1(H^2)}{\mu_0} \nabla
\phi_0 \right ] = -2 \frac{\MU11(H^2)}{\mu_0} \nabla
\phi_0  \cdot (\nabla \phi_0 \cdot \nabla) \nabla \phi_0
\label{eq:phi1lap}
\ea
where
\be
\MU11(x) = \dd{\mu(x)}{x}.
\ee
For a magnetic dipole,
\ba
\phi_0({\bf r}) &=& \frac{{\bf m} \cdot {\bf r}}{|{\bf r}|^3} \\
\nabla \phi_0  \cdot (\nabla \phi_0 \cdot \nabla) \nabla \phi_0 &=& 3
\frac{ \left [ 5 \left ( {\bf m} \cdot {\bf r} \right )^2 + 3 |{\bf
m}|^2 |{\bf r}|^2 \right ] {\bf m}
\cdot {\bf r} }{|{\bf r}|^{13}}
\ea
or more conveniently in spherical coordinates where we have taken the
dipole moment ${\bf m}$ to be aligned along the $z-$axis,
\ba
\phi_0({\bf r}) &=& \sqrt{\frac{4 \pi}{3}} \frac{m}{r^2}
Y_{10}(\theta,\phi)
\label{eq:phi0sph} \\
\nabla \phi_0  \cdot (\nabla \phi_0 \cdot \nabla) \nabla \phi_0 &=& 12
\sqrt{\pi} \frac{m^3}{r^{10}} \left [ \frac{1}{\sqrt{7}} Y_{30} (\phi,\theta) + \sqrt{3}
Y_{10} (\phi,\theta) \right ]
\label{eq:rhoeffsph}
\ea

\section{The Lagrangian to One-Loop Order}

\jcite{Heisenberg and Euler}{Heis36} and \jcite{Weisskopf}{Weis36}
independently derived the effective Lagrangian of the electromagnetic
field using electron-hole theory.  \jcite{Schwinger}{Schw51} later
rederived the same result using quantum electrodynamics.  In
Heaviside-Lorentz units, the Lagrangian is given by
\ba
\lag_0 & = & -{1 \over 4} I \label{eq:lag0def} \\
\lag_1 & = & {e^2 \over h c} \int_0^\infty e^{-\zeta} 
{\d \zeta \over \zeta^3} \left \{ i \zeta^2 {\sqrt{-K} \over 4} \times
\phantom{ 
\cos \left ( {\zeta \over B_k} \sqrt{-{I\over 2} + i \sqrt{K}} \right ) 
\over
\cos \left ( {\zeta \over B_k} \sqrt{-{I\over 2} + i \sqrt{K}} \right ) } 
\right . \\*
\nonumber
& & ~~ \left . 
{ \cos \left ( {\zeta \over B_k} \sqrt{-{I\over 2} + i {\sqrt{-K}\over 2}} \right ) +
\cos \left ( {\zeta \over B_k} \sqrt{-{I\over 2} - i {\sqrt{-K}\over 2}} \right ) \over
\cos \left ( {\zeta \over B_k} \sqrt{-{I\over 2} + i {\sqrt{-K}\over 2}} \right ) -
\cos \left ( {\zeta \over B_k} \sqrt{-{I\over 2} - i {\sqrt{-K}\over 2}} \right ) } 
 + |B_k|^2 + {\zeta^2 \over 6} I \right \}.
\label{eq:lag1def}
\ea
where $B_k = E_k = {m^2 c^3 \over e \hbar} \approx 2.2 \times 10^{15}
\rmmat{\,V\,cm}^{-1} \approx 4.4 \times 10^{13}$\,G.  \jcite{Dittrich
and Reuter}{Ditt85} have derived the second-order corrections to the
Lagrangian and found them to be in general an order of $\alpha$ smaller
than the one-loop corrections regardless of field strength; consequently, 
the one-loop correction should be adequate for all but the most precise 
analyses.

In the weak field limit Heisenberg and Euler \cite{Heis36} give
\be
\lag \approx -{1 \over 4} I + E_k^2 {e^2 \over h c} \left [
{1 \over E_k^4} \left ( {1 \over 180} I^2 - {7 \over 720} K \right
) + {1 \over E_k^6} \left ( {13 \over 5040} K I - {1 \over 630}
I^3 \right ) \cdots \right ]
\label{eq:heweak}
\ee
We define a dimensionless parameter $\xi$ to characterize the field
strength
\be
\xi = {1 \over E_k} \sqrt{I \over 2}
\ee
and use the analytic expression of this Lagrangian for $K=0$
derived by Heyl and Hernquist \cite{Heyl97a}:
\be
\lag_1(I,0) = {e^2 \over h c} \frac{I}{2}
X_0\left(\frac{1}{\xi}\right) \\
\ee
where
\ba
X_0(x)  &=&  4 \int_0^{x/2-1} \ln(\Gamma(v+1)) \d v
+ \frac{1}{3} \ln \left ( \frac{1}{x} \right )
+ 2 \ln 4\pi - (4 \ln A+\frac{5}{3} \ln 2) \nonumber \\
& & ~~ - \left [ \ln 4\pi + 1 +  \ln \left ( \frac{1}{x} \right ) \right ] x
+ \left [ \frac{3}{4} + \frac{1}{2} \ln \left ( \frac{2}{x} \right )
\right ]
x^2
\label{eq:x0anal} 
\ea
where
\be
\ln A = \frac{1}{12} - \zeta^{(1)}(-1).
\ee
Here $\zeta^{(1)}(x)$ denotes the first derivative of the Riemann Zeta
function.

With the analytic form of the Lagrangian, calculating $\mu'_0$ and
$\mu'_1$ is straightforward and $\alpha$ provides a convenient
ordering parameter.
\ba
\mu'_0 &=& -4 \pp{\lag_0}{I} = 1 \\
\mu'_1 &=& -\frac{\alpha}{2 \pi}
 \left[2 X_0 \left(\frac{1}{\xi}\right) - \frac{1}{\xi} X_0^{(1)}
\left(\frac{1}{\xi}\right) \right ] 
\ea
where $\xi=B/B_k$, and
\be
X_0^{(1)}(x) = \dd{X_0(x)}{x}.
\ee
Inverting this relationship to first order yields,
\ba
\mu_0 &=& 1 \\
\mu_1(H^2) &=& \frac{\alpha}{2 \pi}
\left[2 X_0 \left(\frac{B_k}{H}\right) - \frac{B_k}{H} X_0^{(1)}
\left(\frac{B_k}{H}\right) \right ] \\
\MU11(H^2) &=&
\frac{\alpha}{2 \pi} \frac{1}{2 B_k^2}
\left[\frac{B_k^4}{H^4} X_0^{(2)} \left(\frac{B_k}{H}\right) -
\frac{B_k^3}{H^3} X_0^{(1)} \left(\frac{B_k}{H}\right) \right ] \\
 & = & \frac{\alpha}{2 \pi} \frac{1}{6 B_k^2} \left \{ 2
\frac{B_k^2}{H^2} + 3 \frac{B_k^3}{H^3} \left [ \ln (4\pi) + 1 + \ln
\left ( \frac{H}{B_k} \right ) - 2 \ln \Gamma \left ( \frac{1}{2} \frac{B_k}{H}
\right ) \right ] + 3 \frac{B_k^4}{H^4} \left [ \psi \left(
\frac{1}{2} \frac{B_k}{H} \right ) - 1 \right ] \right \}  
\label{eq:mu11gen}
\ea
where $\psi(x)$ is the digamma function,
\be
\psi(x) = \dd{\ln \Gamma (x)}{x}.
\ee
The expression for $\mu_1$ agrees numerically with the results of
Mielnieczuk \etal \cite{Miel88}.  The function $\MU11(H^2)$ may
conveniently be expanded in the weak-field limit ($H<B_k/2$),
\be
\MU11(H^2) = -\frac{\alpha}{2\pi} \frac{8}{B_k^2}
\sum_{j=0}^\infty \frac{2^{2j} B_{2(j+2)} }{2j+3}
\left(\frac{H}{B_k}\right)^{2j}
\label{eq:mu11weak}
\ee
where $B_j$ denotes the $j$th Bernoulli number, 
and in the strong-field limit ($H>B_k/2$)
\be
\MU11(H^2) = \frac{\alpha}{2\pi} \frac{1}{B_k^2} \left \{ \frac{1}{3} \frac{B_k^2}{H^2} - \frac{1}{2}
\frac{B_k^3}{H^3} \left [ \ln \left ( \frac{H}{B_k} \right ) + 1 -
\ln\pi \right ] - \frac{1}{2} \frac{B_k^4}{H^4} - \sum_{j=5}^\infty
\frac{(-1)^j}{2^{j-3}} \frac{j-4}{j-3} \zeta (j-3) \left (
\frac{H}{B_k} \right )^{-j}   \right \}.
\label{eq:mu11strong}
\ee
where we have used the expansions of Ref.~\cite{Heyl97a}.

\begin{figure}
\plotonesc{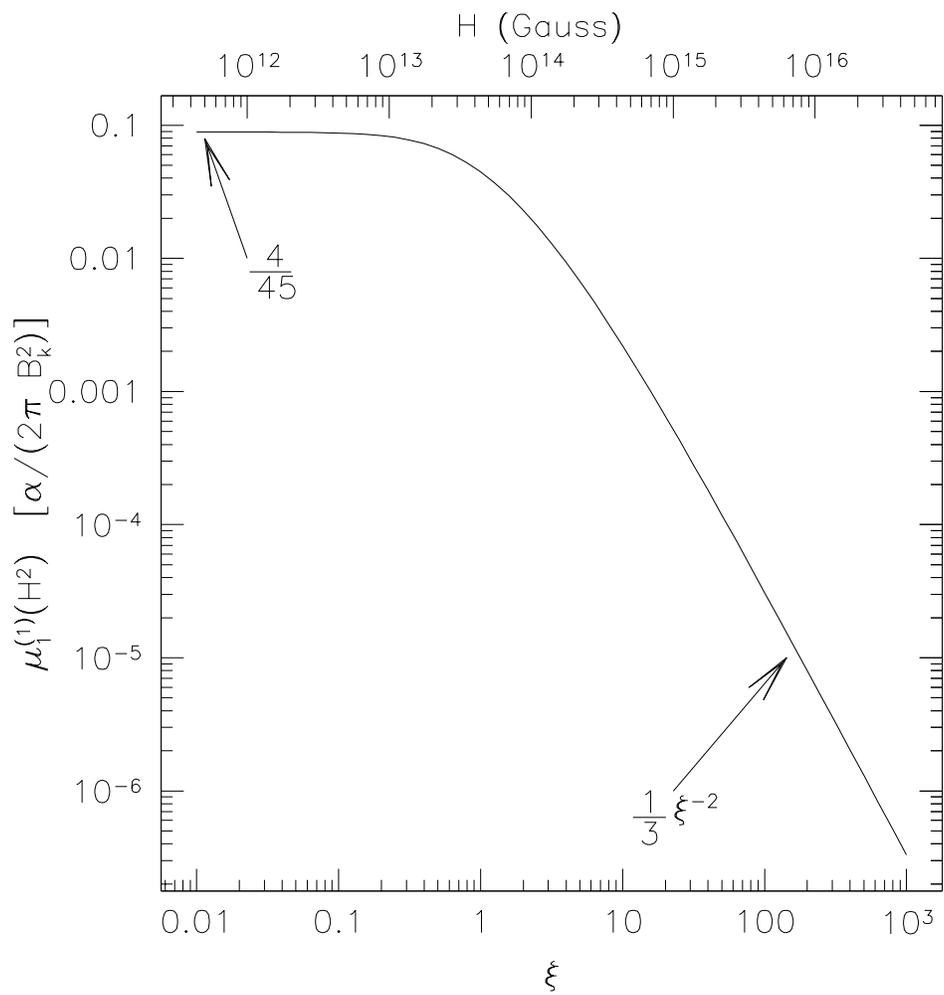}{0.7}
\caption{$\MU11(H^2)$ as a function of $\xi, H$}
\label{fig:mu11}
\end{figure}

As apparent from \figref{mu11}, $\MU11(H^2)$ is constant up to
approximately $H=0.5 B_k$ and then begins to decrease quickly as
$H^{-2}$.   The existence of these two regimes allows us to find
analytic solutions for the correction to the potential ($\phi_1$).

\section{Solving for the first-order correction}

\subsection{Weak-field limit}

$\MU11(H^2)$ is constant as long as $H \ll B_k$.  In this regime
\eqref{phi1lap} may be solved analytically.  Since spherical harmonics
are eigenfunctions of the angular component of the Laplacian operator,
it is expedient to expand the right-hand side of \eqref{phi1lap} in
terms of spherical harmonics \cite{Binn87}.  
\be
\rho_{lm}(r) = \int_0^\pi \sin \theta \d \theta \int_0^{2\pi} \d \phi
Y_{lm}^*(\theta,\phi) \RHOeff(r,\theta,\phi)
\label{eq:rhoexp}
\ee
With this definition the correction to the potential is given by
\cite{Binn87} 
\be
\phi_1(r,\theta,\phi) = - \sum_{l,m} \frac{Y_{lm}(\theta,\phi)}{2l+1}
\left [ \frac{1}{r^{l+1}} \int_0^r \rho_{lm}(a) a^{l+2} \d a
+ r^l \int_r^\infty \rho_{lm}(a) \frac{\d a}{a^{l-1}} \right ]
\label{eq:phi1sol}
\ee
If $\MU11(H^2)$ is a constant, we see from \eqref{phi1lap} and
\eqref{rhoeffsph} that the expansion given by \eqref{rhoexp} is
straightforward
\ba
\rho_{10}(r) &=& -24 \sqrt{3 \pi} \frac{\MU11}{\mu_0} \frac{m^3}{r^{10}}
\label{eq:rho10} \\
\rho_{30}(r) &=& -24 \sqrt{\frac{\pi}{7}}
\frac{\MU11}{\mu_0} \frac{m^3}{r^{10}} \label{eq:rho30}
\ea

If we combine these results with \eqref{phi1sol} we see that the interior
integral diverges if $\MU11$ is a constant.  This does not present a
problem if we insert an inner bound ($r_0$) to the interior integral.  This
inner bound is defined as the radius at which either $\MU11(H^2)$
begins to change (\ie when $H \agt 0.5 B_k$) or when the zeroth order
potential is no longer given by the dipole formula (\ie at the surface
of the object). We obtain
\ba
\phi_{1,10}(r,\theta,\phi) &=& \frac{4}{9} \sqrt{3\pi}
\frac{\MU11}{\mu_0} \frac{m^3}{r^2} \left ( \frac{3}{r_0^6} -
\frac{1}{r^6} \right ) Y_{10} (\theta,\phi) \\
\phi_{1,30}(r,\theta,\phi) &=& 6 \sqrt{\frac{\pi}{7}} \frac{\MU11}{\mu_0}
\frac{m^3}{r^4}  \left ( \frac{1}{7} \frac{1}{r_0^4} - \frac{1}{11}
\frac{1}{r^4} \right ) Y_{30} (\theta,\phi).
\ea
These functions describe radially dependent corrections to the dipole
and hexapole moments of the object under consideration.  If we define
the higher moments in analogy to \eqref{phi0sph},
\be
\phi_{l0} (r,\theta,\phi) = \sqrt{\frac{4\pi}{2l+1}} \frac{M_{l0}}{r^{(l+1)}}
Y_{l0} (\theta,\phi),
\ee
we obtain the following corrections
\ba
m_1(r) &=& \frac{2}{3} \frac{\MU11}{\mu_0} m^3 \left (
\frac{3}{r_0^6} - \frac{1}{r^6} \right ) \\
M_{30,1}(r) &=& 3 \frac{\MU11}{\mu_0} m^3 \left (
\frac{1}{7} \frac{1}{r_0^4} - \frac{1}{11} \frac{1}{r^4} \right ) .
\ea
To eliminate the dependence on the inner bound, $r_0$, we assume that
the dipole and hexapole moments are known at a radius $r_s > r_0$ and
calculate the difference between the known moments at $r_s$ and those
measured at infinity.

Substituting the values of $\mu_0$ and $\MU11$, we obtain,
\ba
m(r=\infty) - m(r=r_s) &=&
\Delta m = \frac{8}{135} \frac{\alpha}{2 \pi} m \left ( \frac{m}{r_s^3}
\frac{1}{B_k} \right )^2 \\
M_{30}(r=\infty) - M_{30}(r=r_s) &=& \Delta M_{30} = \frac{4}{165}
\frac{\alpha}{2\pi} m r_s^2 \left ( \frac{m}{r_s^3} \frac{1}{B_k}
\right )^2.
\ea
Both the corrections are given in terms of the magnetic field strength
at $r_s$.  Since we have assumed that $\MU11(H^2)$ is constant near
$r_s$, the field strength at $r_s$ must appreciably be less than
$B_k$ for this set of approximations to be valid.  Consequently, the
correction to the dipole moment is indeed quite small, less than
one-thousandth of the ``bare'' dipole moment.  However, there is a
correction to the hexapole moment even without a ``bare'' hexapole
moment.  Thus radiative corrections generate a hexapole field
which is in principle measurable at infinity.

\subsection{The General Case}

In the strong-field limit, we must use the general equation
(\eqref{mu11gen}) for $\MU11(H^2)$ because the magnetic field
strength varies as a function of $\theta$ around the dipole.  Before
tackling this problem numerically, we can glean several
characteristics of the solutions from \eqref{phi1lap} and
\eqref{rhoexp}.  Since for a dipole $\RHOeff$ is a odd function of
$\theta$ and constant with respect to $\phi$,
the contributions,
\be
\rho_{lm}(r) = 0 \rmmat{~if~} l \rmmat{~is even or~} m \neq 0.
\ee
Furthermore, we can more conveniently write
\be
\rho_{lm}(r) = \frac{1}{\mu_0} \frac{\alpha}{2\pi} \frac{1}{B_k^2}
\frac{m^3}{r^{10}} \chi_{lm} (\beta)
\ee
where $\chi_{lm}(\beta)$ is a dimensionless function of a
dimensionless argument,
\be
\beta =\frac{m}{r^3} \frac{1}{B_k}.
\ee
For a dipole,
\be
\chi_{lm}(\beta) = -2 \int_0^\pi \sin \theta \d \theta \int_0^{2\pi} \d \phi
Y_{lm}^*(\theta,\phi) \left ( \frac{\alpha}{2\pi} \frac{1}{B_k^2}
\right)^{-1}
\Biggr ( 3 \cos \theta \left ( 5 \cos^2
\theta +3 \right ) \Biggr )
\MU11 \left [ \beta^2 B_k^2 \left ( 3 \cos^2
\theta + 1 \right ) \right ] .
\label{eq:rhodim}
\ee

\begin{figure}
\plotonesc{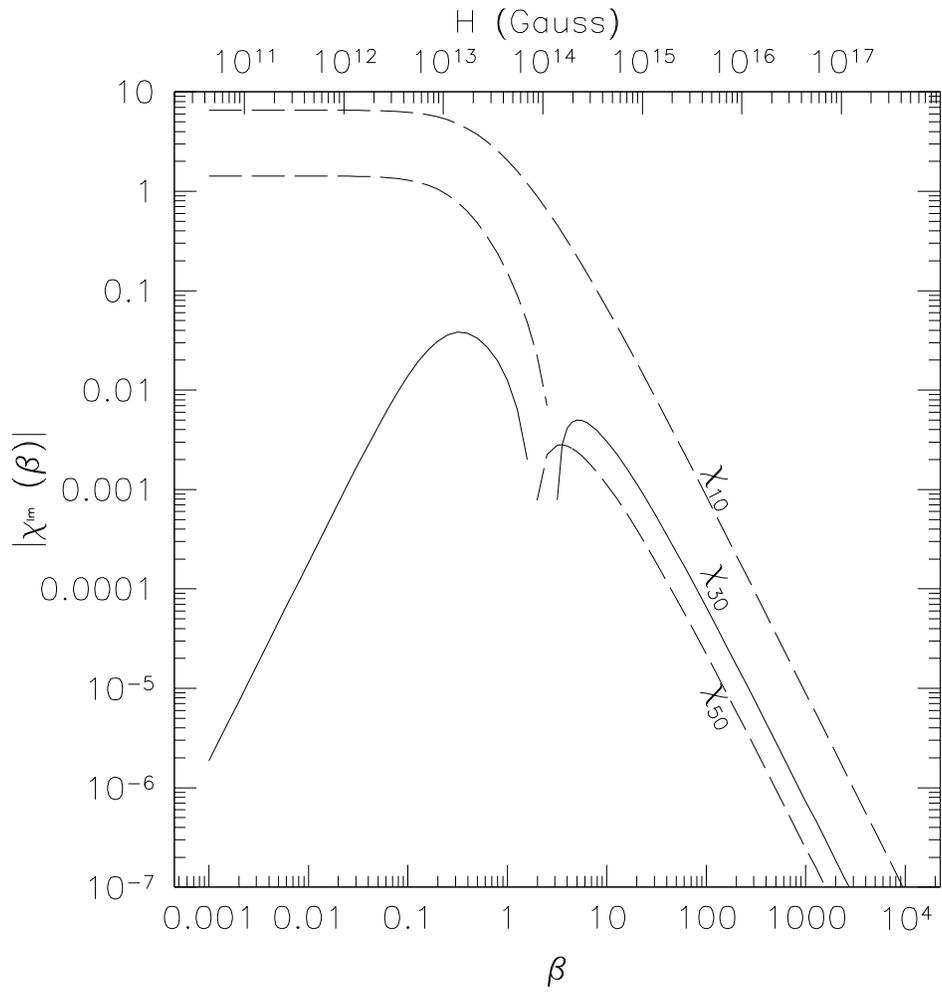}{0.7}
\caption{$\chi_{lm}(\beta)$ as a function of $\beta, H$.  The solid
lines indicate where $\chi_{lm}(\beta)$ is positive.  The dashed
lines indicate negative values of the ordinate.}
\label{fig:varrho}
\end{figure}

We calculate numerically the functions $\chi_{lm}(\beta)$ for the
first three odd harmonics and depict the results in \figref{varrho}.

The limiting expressions for the weak-field
limit are easily calculated,
\ba
\chi_{10}(\beta) &=& \sqrt{3\pi} \left [ -\frac{32}{15} + \frac{13312}{1225}
\beta^2 - \frac{167936}{1225} \beta^4 + {\cal O} \left (\beta^6
\right) \right ] \\
\chi_{30}(\beta) &=& \sqrt{7\pi} \left [ -\frac{32}{105} + \frac{11776}{3675}
\beta^2 - \frac{323584}{5775} \beta^4 + {\cal O} \left (\beta^6
\right) \right ] \\
\chi_{50}(\beta) &=& \sqrt{11\pi} \left [ +\frac{512}{1617} \beta^2
- \frac{1150976}{105105} \beta^4 + {\cal O} \left (\beta^6
\right) \right ] 
\ea
In the strong-field limit, we have the following approximations
\ba
\chi_{10}(\beta) &=& -\frac{4}{27} \sqrt{\pi} \left (27 \sqrt{3} - 4
\pi \right) \beta^{-2} + {\cal O} \left ( \beta^{-3} \right ) \\
\chi_{30}(\beta) &=& \frac{8}{243} \sqrt{7\pi} \left (81 -
14\sqrt{3}\pi \right) \beta^{-2} + {\cal O} \left ( \beta^{-3} \right ) \\
\chi_{50}(\beta) &=& -\frac{4}{1215} \sqrt{11\pi} \left (1863 - 340 \sqrt{3}
\pi \right) \beta^{-2} + {\cal O} \left ( \beta^{-3} \right ) 
\ea
The integrals for each spherical-harmonic component of
the first-order correction (\eqref{phi1sol}) may be recast
in terms of integrals over $\beta$
\be
\phi_1(r,\theta,\phi) = -\frac{1}{3} \frac{m}{r^2}  \frac{1}{\mu_0}
\frac{\alpha}{2\pi} \beta^2
\sum_{l,m} \frac{Y_{lm}(\theta,\phi)}{2l+1}
\left [ \beta^{(l-7)/3} \int_\beta^{\beta_0} \chi_{lm}(\upsilon) \upsilon^{-(l-4)/3} \d \upsilon
+ \beta^{-(l+8)/3} \int_0^\beta \chi_{lm}(\upsilon)
\upsilon^{(l+5)/3} \d \upsilon \right ]
\label{eq:phi1strong}
\ee
where the first integral is to be evaluated in the limit as $\beta_0
\rarrow \infty$.  Although \eqref{phi1strong} appears to be scale
free, the cutoff $\beta_0$ has a physical interpretation.  
Firstly, it can be taken to be the magnetic field strength at the surface of
the object.  For a point magnetic dipole (\eg an electron), the
interpretation is more subtle.  As one approaches a point dipole, not only
does the field become arbitrarily strong, so do the field gradients.
When these gradients become larger in magnitude than $B_k/\lambda_e$
($\lambda_e=\hbar/m_e c$, the electron Compton wavelength),
the Heisenberg-Euler Lagrangian is no longer applicable; therefore,
we do not expect our expressions for $\RHOeff$ to be valid arbitrarily
close to a point dipole.  The radius or field strength at which our
expression for $\RHOeff$ fails depends on the intrinsic dipole moment
of the object ($m$),
\be
r_0 \sim \left ( \frac{m}{B_k} \lambda_e \right )^{1/4}
\rmmat{~~~or~~~}
\beta_0 \sim \left (\frac{m}{B_k}\right)^{1/4} \lambda_e^{-3/4}
\ee

\begin{figure}
\plotonesc{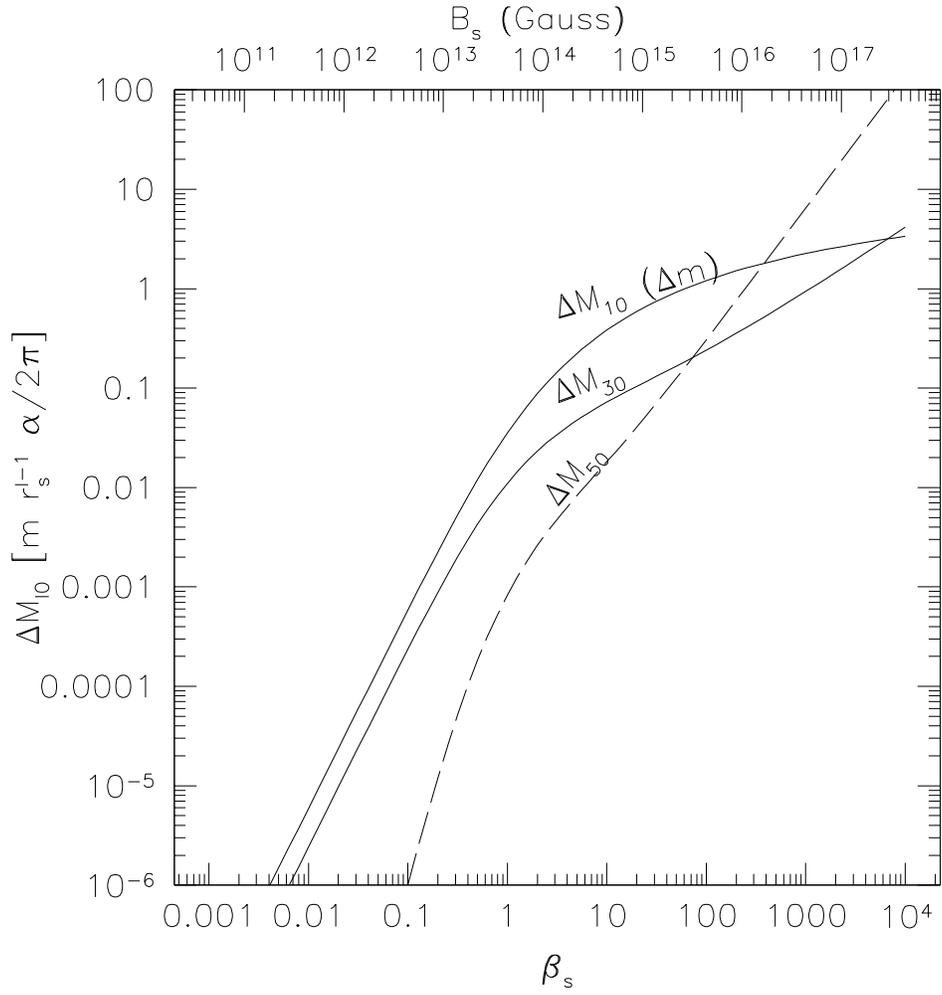}{0.7}
\caption{The difference in the strength of the multipole moments
measured at infinity ($\beta\rarrow 0$) and at the surface, $M_{l0}(0)-M_{l0}(\beta_s)$ ($\Delta M_{l0}$), as a function
the strength of the dipole field at the surface, $\beta_s, H_s$.
The solid lines indicate the moments for which
the vacuum acts paramagnetically, \ie
$\Delta M_{l0}(\beta_s)>0$.
The dashed lines indicate negative values of the ordinate.}
\label{fig:cumrho}
\end{figure}

As in the weak-field case, we calculate the shift in the observed
multipole moments at infinity relative to the known moments at some
inner radius $r_s>r_0$ or equivalently at some field strength
$\beta_s<\beta_0$.
\ba
m_1(0)-m_1(\beta_s) &=& -\frac{1}{9} m \frac{1}{\mu_0}
\frac{\alpha}{2\pi} \sqrt{\frac{3}{4\pi}} \left [ \int_0^{\beta_s}
\chi_{10}(\upsilon) \upsilon \d \upsilon - \frac{1}{\beta_s}
\int_0^{\beta_s} \chi_{10}(\upsilon) \upsilon^2 \d \upsilon \right
] \\
M_{30,1}(0)-M_{30,1}(\beta_s) &=& -\frac{1}{21} m r_s^2 \frac{1}{\mu_0}
\frac{\alpha}{2\pi} \sqrt{\frac{7}{4\pi}} \left [ \beta_s^{2/3} \int_0^{\beta_s}
\chi_{30}(\upsilon) \upsilon^{1/3} \d \upsilon - \beta_s^{-5/3}
\int_0^{\beta_s} \chi_{30}(\upsilon) \upsilon^{8/3} \d \upsilon \right
] \\
M_{50,1}(0)-M_{50,1}(\beta_s) &=& -\frac{1}{33} m r_s^4 \frac{1}{\mu_0}
\frac{\alpha}{2\pi} \sqrt{\frac{11}{4\pi}} \left [ \beta_s^{4/3} \int_0^{\beta_s}
\chi_{50}(\upsilon) \upsilon^{-1/3} \d \upsilon - \beta_s^{-7/3}
\int_0^{\beta_s} \chi_{50}(\upsilon) \upsilon^{10/3} \d \upsilon \right
] 
\ea
and in general
\ba
M_{l0,1}(0)-M_{l0,1}(\beta_s) &=& -\frac{1}{3(2l+1)} m r_s^{l-1} \frac{1}{\mu_0}
\frac{\alpha}{2\pi} \sqrt{\frac{2l+1}{4\pi}} \Biggr [ \beta_s^{(l-1)/3} \int_0^{\beta_s}
\chi_{l0}(\upsilon) \upsilon^{-(l-4)/3} \d \upsilon \nonumber \\*
 & & ~~~
- \beta_s^{-(l+2)/3}
\int_0^{\beta_s} \chi_{l0}(\upsilon) \upsilon^{(l+5)/3} \d \upsilon \Biggr
] 
\ea
\figref{cumrho} depicts the shifts in the moments between the
surface of the dipole and infinity.

For the higher moments ($l>1$), the integrals in \eqref{phi1strong}
are well behaved so we need not set a inner bound ($\beta_0$) and we
can define in general
\be
M_{l0,1}(0) = -\frac{1}{3(2l+1)} m \left (\frac{m}{B_k} \right
)^{(l-1)/3} \frac{1}{\mu_0}
\frac{\alpha}{2\pi} \sqrt{\frac{2l+1}{4\pi}} \int_0^\infty
\chi_{l0}(\upsilon) \upsilon^{-(l-4)/3} \d \upsilon 
\ee
This integral converges for all $l>1$.  For $l=1$, it diverges
logarithmically as $\upsilon\rarrow\infty$.

\section{Conclusions: Application to Neutron Stars} 

The environment of a neutron star is strongly magnetized; therefore, the
one-loop QED corrections may be significant, especially for those
neutron stars with ultrastrong surface fields (magnetars) \cite{Dunc92}.
Although plasma probably fills this region, QED vacuum corrections
dominate the contribution of the plasma to the magnetic permeability
\cite{Mesz92}. 

The structure of the magnetic field at the surface of a neutron star
is an important clue to the origin of neutron-star magnetic fields.
Several authors have proposed \cite{Blan83} that currents in the thin
crust generate the observed magnetic fields.  In this case, the field
structure will be dominated by high-order multipoles with $l \sim
\delta r_c/r_* \gg 1$ where $\delta r_c$ is the thickness of the crust
and $r_*$ is the radius of the star \cite{Aron93}.  Arons
\cite{Aron93} argues further that the observed spin-up line for
millisecond pulsars constrains the strength of higher-order multipoles
at the surface to be no more than 40 \% of surface strength of the
dipole.

\begin{figure}
\plotonesc{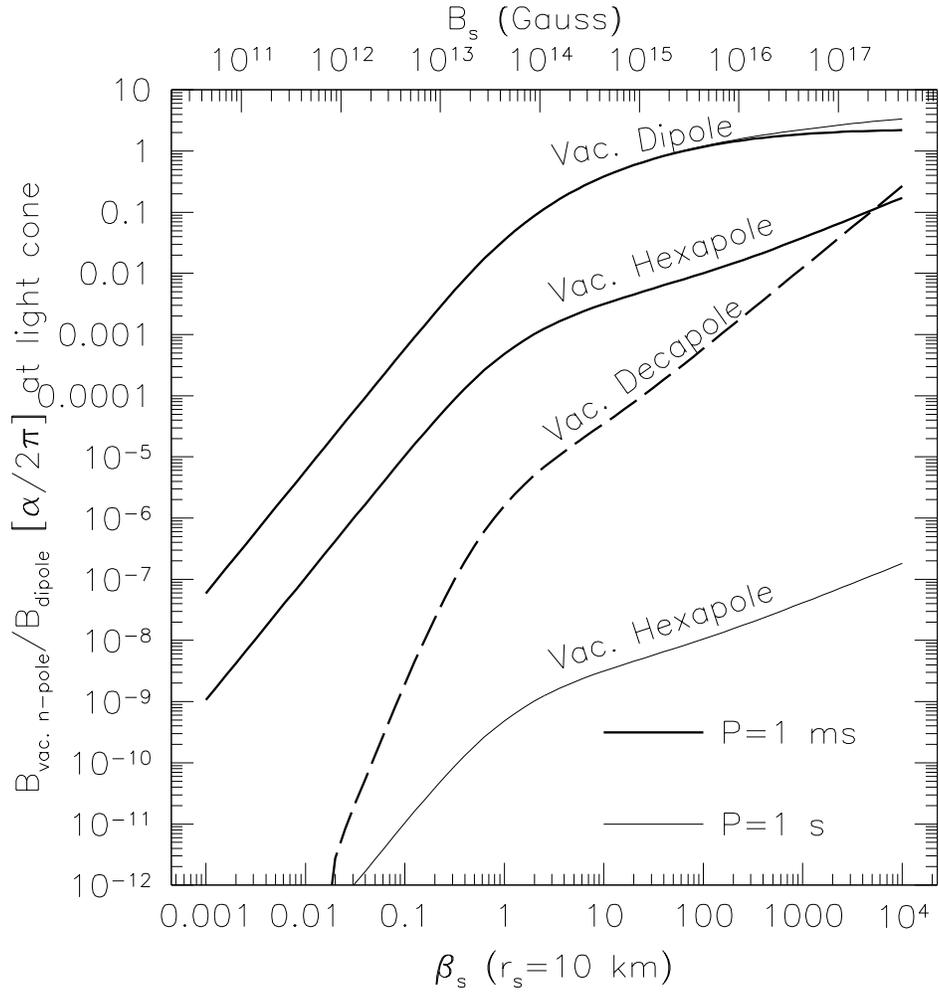}{0.7}
\caption{The fractional contribution of higher magnetic moments to the
field strength at the
light cylinder as function of surface dipole field strength for
periods of one millisecond and one second.  The contribution of the
vacuum decapole at the light cone for $P=1$s is too small to be
depicted on this graph.}
\label{fig:cperc}
\end{figure}

The current results complicate this argument.  The location of the
observed spin-up line and the value of spin-down index of a pulsar
depend on the strength of the various moments of the magnetic field at
the light cylinder.  Our results show that the vacuum itself may
generate higher magnetic moments between the neutron star surface and
the light cylinder.   \figref{cperc} depicts the fractional
contribution of higher magnetic moments to the field strength at the
light cylinder as a
function of the surface dipole field strength and the pulsar period.

Even for magnetars near their birth, the vacuum adds only a small
correction to the field strength at the light cone.  Because of the
weakness of the QED coupling, the one-loop corrections to otherwise
classical descriptions of a magnetic dipole tend to be small for all
but the most extreme field strengths.

Observing this effect would be difficult.  For a magnetar,
measurements of the field strength at two different radii and an
estimate of the strength of higher order moments at the surface each
to a precision of one part in one-thousand would be required.
However, if one could argue that an object had no hexapole
or higher-order multipole intrinsically, a measurement of a
higher-order multipole far from the object would uncover the effects
of one-loop corrections.  For example, an electron is intrinsically a
magnetic dipole.  If one ignores the contribution of terms in the
Lagrangian depending on field gradients, one would expect the vacuum
surrounding an electron to generate higher order multipole fields.
Determining whether this occurs is beyond the scope of this work.

\end{document}